\documentclass[review]{elsarticle}

\usepackage{hyperref}
\usepackage{bm}
\usepackage{multirow}
\usepackage{amsmath}
\usepackage[caption=false]{subfig}

\begin{document}

\begin{frontmatter}

\title{Role of shape on the forces on an intruder moving through a dense granular medium}


\author{Bitang Kwrung Tripura,\textit{$^{a}$} Sonu Kumar, \textit{$^{a}$} K. Anki Reddy, \textit{$^{a}$} Julian Talbot \textit{$^{b}$}}
\address{\textit{$^{a}$} Department of Chemical Engineering, Indian Institute of Technology, Guwahati, Assam-781039, India. E-mail: anki.reddy@iitg.ac.in, Telephone: +91-361-258-3532}
\address{\textit{$^{b}$} Sorbonne Universite, CNRS, Laboratoire de Physique Theorique de la Matiere Condensee, LPTMC, F-75005 Paris, France}



\begin{abstract}

We use numerical simulation  to investigate the effect an intruder's shape has on the drag and lift forces  that it experiences while moving through a granular medium composed of polydisperse disks of mean diameter $d$. The intruder velocity, $v$, was varied from 0.1$\sqrt{dg}$ to 20$\sqrt{dg}$. For frictionless particles ($\mu=0.0$) there is a gradual increase in drag force with increasing $v$, whereas for frictional systems ($\mu=0.1, 0.5$) a constant drag regime appears at low velocities. The drag force depends weakly on the shape of the object provided that the cross section is same. The drag force depends linearly on the immersion depth, while there is very little variation in lift force with depth for certain shapes. The lift experienced by the object is a strong function of its shape at a given velocity. Shape has an effect on the distribution of contacts around the surface of the intruder which may result in a strong lift for certain shapes. We also show the force profiles around the intruder surface.

\end{abstract}


\end{frontmatter}



\section{Introduction}
Granular matter is a collection of discrete macroscopic particles that exhibit properties of solids or fluids depending on the volume fraction\cite{jaeger}. It is present everywhere around us in multiple forms such as sand in the deserts or as grains in food industries. Dry granular matter exhibits properties that resemble those of Newtonian fluids such as capillary action\cite{capillary}, Magnus effect\cite{magnus}, Kevin-Helmholtz\cite{kevinhelm} and Rayleigh-Taylor\cite{rayltay} instabilities. In addition, understanding the forces on objects moving in fluids has been an active area of research for the last few decades, as it has applications in the development of vehicles in automobile industry\cite{huminic2017aerodynamic} and  aerospace industry\cite{bushnell2003aircraft}, etc. Though drag and lift forces  are well understood in fluids but they  have been less  investigated in granular media.

Drag is the retarding force exerted on a body by surrounding medium. Studies  performed  for various configurations\cite{slow_drag,albert,draginduced} in granular media to understand the drag include an intruder object  moving horizontally or vertically  at low \cite{slow_drag} or high velocities \cite{potiguar2013lift}. In the slow velocity regime the drag force is either independent or weakly dependent of the intruder's velocity\cite{slow_drag}. At higher velocity the drag force increases monotonically with the intruder velocity. Albert \textit{et al.} \cite{albert}  studied objects of different shapes and found that the difference in drag for any pair was less than twenty percent in the low-velocity regime. This work, however, did not explore the behavior at higher velocities. In a recent study of intruders of various shapes in a slow granular silo flow it was observed that the dimensionless number characterizing the drag force varied significantly with shape \cite{katsu}.

In addition to drag, objects may also experience a lift force while moving in a granular medium. However, there have been very few studies devoted to understand this phenomenon \cite{draginduced,potiguar2011lift,stress3,debnath2017lift}. Ding \textit{et al.}\cite{draginduced}  stated that a symmetric object, such as a cylinder, when dragged within a granular medium  experiences a weak lift force that varies linearly with the depth. Potiguar \cite{potiguar2011lift} reported that in a dilute granular flow there was no net lift force  on a circular obstacle, while the net lift force on an asymmetric obstacle depends non-monotonically on the intruder velocity and depth. Guillard \textit{et al.}\cite{stress3} studied the lift force on a cylinder moving horizontally in a granular medium under gravity. They observed that the lift force saturates at depths greater than the cylinder diameter and noted that the gravitational pressure gradient breaks the up/down symmetry of the moving intruder thereby modifying the flow around its surface. Debnath \textit{et al.}\cite{debnath2017lift} observed that the lift force for  a disc-shaped intruder rises with the immersion depth and reaches a constant value at larger depths. It was argued that the lift is the result of the asymmetry in the dilation and shear rate in the regions above and below the intruder.

The studies mentioned above are relevant for understanding animal locomotion in granular media. Subsurface motion is essential for sand dwelling animals to shelter from high temperatures in deserts during day time or to escape from the predators. Animals such as sand lizards propel themselves using undulatory movements \cite{maladen,ding}. Hence, it is crucial to address the effect of shape on drag, and the induced lift as the shape of the object determines the strength of the jammed region built in front of the intruder\cite{albert}.

The objective of the present work is to understand the  shape and friction dependence of the drag and lift forces  on an intruder immersed  in a granular medium as a function of its velocity and depth. The paper is organized as follows.  We provide details about the simulation method  in section~\ref{sec:simumethod}, results and discussion in section~\ref{sec:resdis}, followed by our conclusions in section~\ref{sec:conclu}.

\section{Simulation Methodology}\label{sec:simumethod}
In this work, we employed the Discrete Element Method (DEM)\cite{cundall} to study the forces on variously-shaped objects being dragged through a granular medium in the presence of gravity. The simulations were performed in two dimensions with periodic boundary conditions in the $x$-direction. A  gravitational field of magnitude $g$ acts along the negative $y$ direction and a wall, composed of particles of diameter $d$,  confines the simulation system along $y=0$. We poured  randomly 63000 particles, of diameters uniformly distributed in the range $0.9-1.1d$, from above after placing an intruder in the system. The particles were allowed to settle under the influence of gravitational and dissipative forces (discussed later) until the energy of the system reached a minimum. The simulation system spans $300d$ along the $x$ direction while the free surface is present at a height of about $195d$ measured from the bottom. The depth of the center of mass of the intruder from this free surface is denoted by $h$. The density (mass per unit area) is set as $\rho$.
left bot right top
\begin{figure}
      \includegraphics[clip,trim={4cm 10.0cm 0.5cm 14cm}, width=0.95\linewidth]{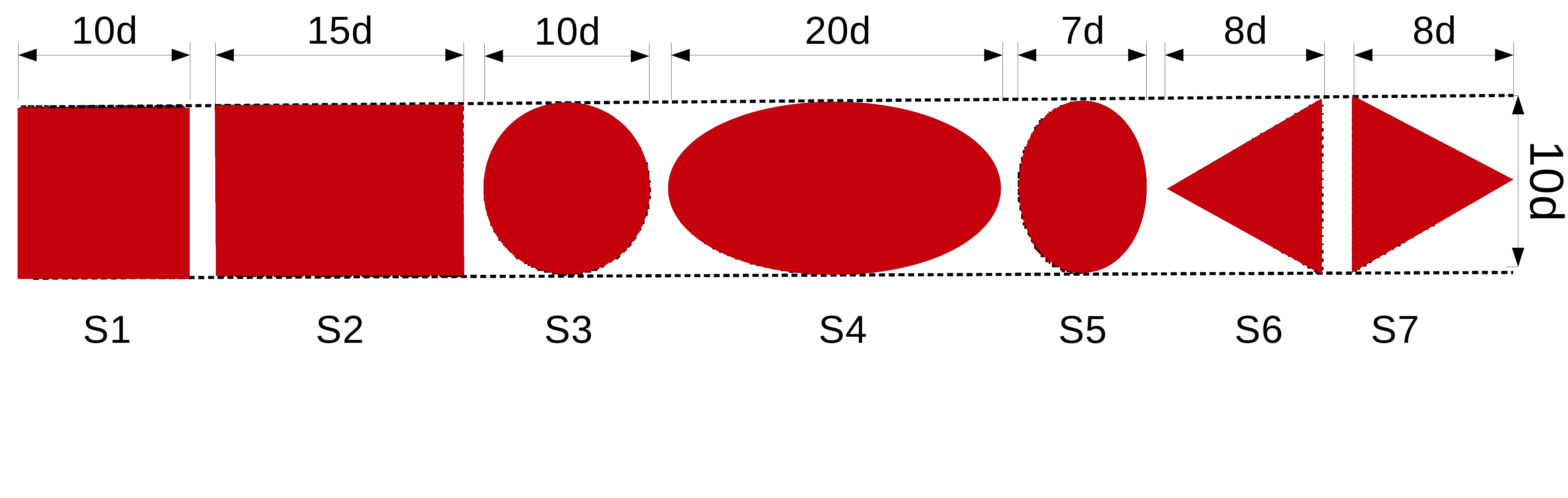}
        \caption{The various shapes considered in our study. In the simulations they move from left to right.}
                \label{fig:shapes}
\end{figure}

\begin{figure}
        \includegraphics[width=0.9\linewidth]{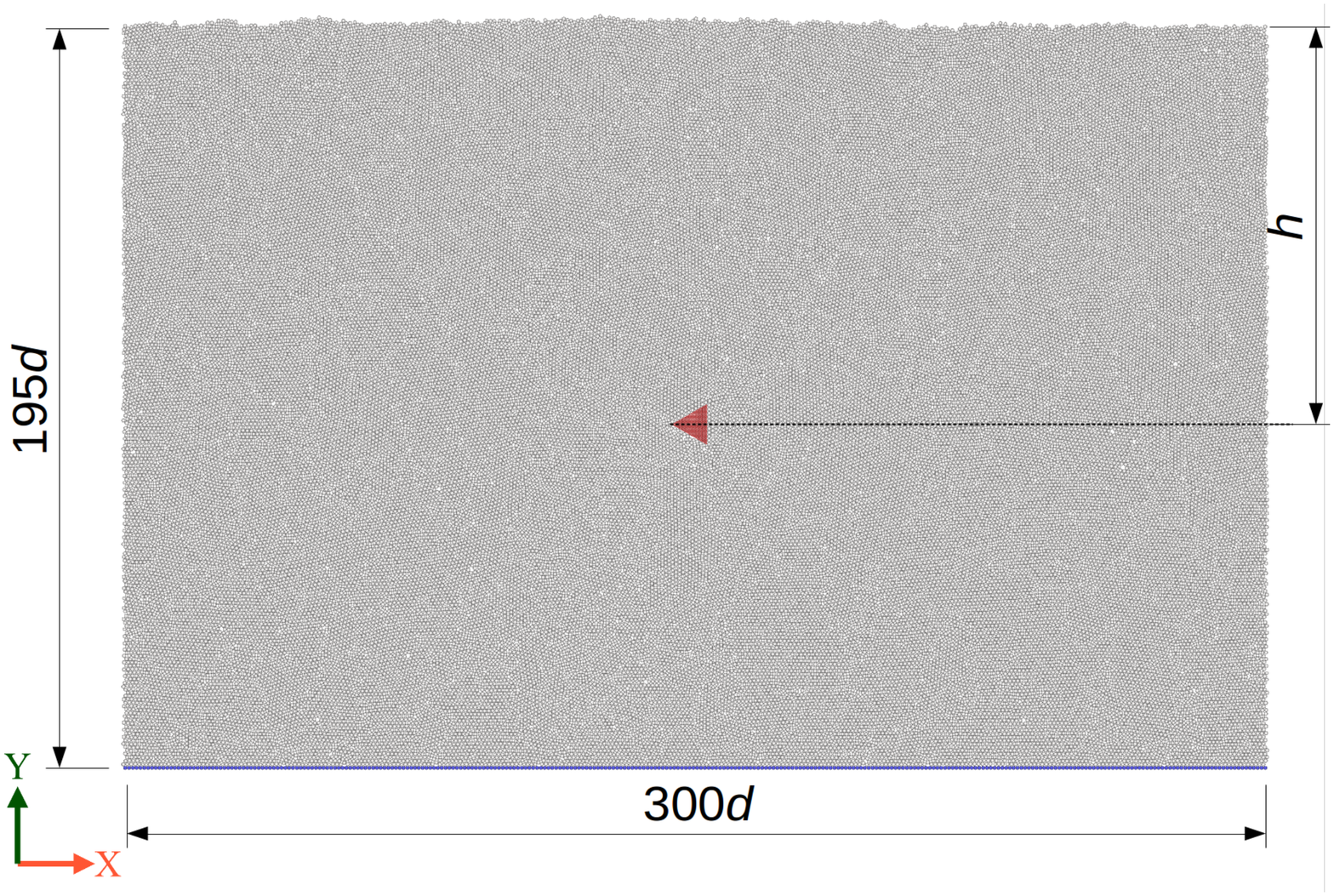}

        \caption{Initial configuration for one of the shapes (backward triangle, S6) at a depth $h=105d$. Periodic boundary conditions are applied in the $x$-direction. The system is confined by a wall at $y=0$ composed of particles of size $1d$(blue particles) while the top surface is unconstrained. The origin is located at the left bottom corner.}
        \label{fig:init}
\end{figure}

The intruder is displaced at a constant velocity $v$ along the positive $x$-direction for a total distance of 1200$d$. We considered 7 different shapes in our study: a square (S1), a rectangle (S2), a disk (S3), an ellipse with major axis aligned with the direction of motion of the intruder (S4), an ellipse with minor axis aligned with the x- direction (S5), an equilateral triangle with edge pointing opposite to the moving direction (S6) and an equilateral triangle with edge pointing along the moving direction (S7): See Fig. \ref{fig:shapes}. The cross-section facing the flow  of all shapes has the same length of $10d$. Fig. \ref{fig:init} shows an initial configuration for S6 at a depth of $h=105d$.

To update the positions and velocities of particles as a function of space and time, one needs to compute the forces they experience due to interaction with the neighboring particles. The normal $(\boldsymbol{f_{ij}^n})$ and tangential $(\boldsymbol{f_{ij}^t})$ contact forces between particles $i$ and $j$ were calculated with the following expressions \cite{Brilliantov,silbert2001}.

\begin{equation}
\boldsymbol{f_{ij}^n}=\sqrt{\frac{d_i d_j}{2(d_i+d_j)}}\sqrt{\xi_{ij}}(k_n \xi_{ij} \boldsymbol{\hat{n}_{ij}}-m_{\textrm{eff}}\gamma_n \boldsymbol{\dot{r}_{ij}^n})
\end{equation}

\begin{equation}
\boldsymbol{f_{ij}^t}=-\sqrt{\frac{d_i d_j}{2(d_i+d_j)}}\sqrt{\xi_{ij}}(k_t \boldsymbol{\Delta s_{ij}}+m_{\textrm{eff}}\gamma_t \boldsymbol{\dot{r}_{ij}^t)}
\end{equation}

\begin{figure*}
        \includegraphics[width=0.9\linewidth]{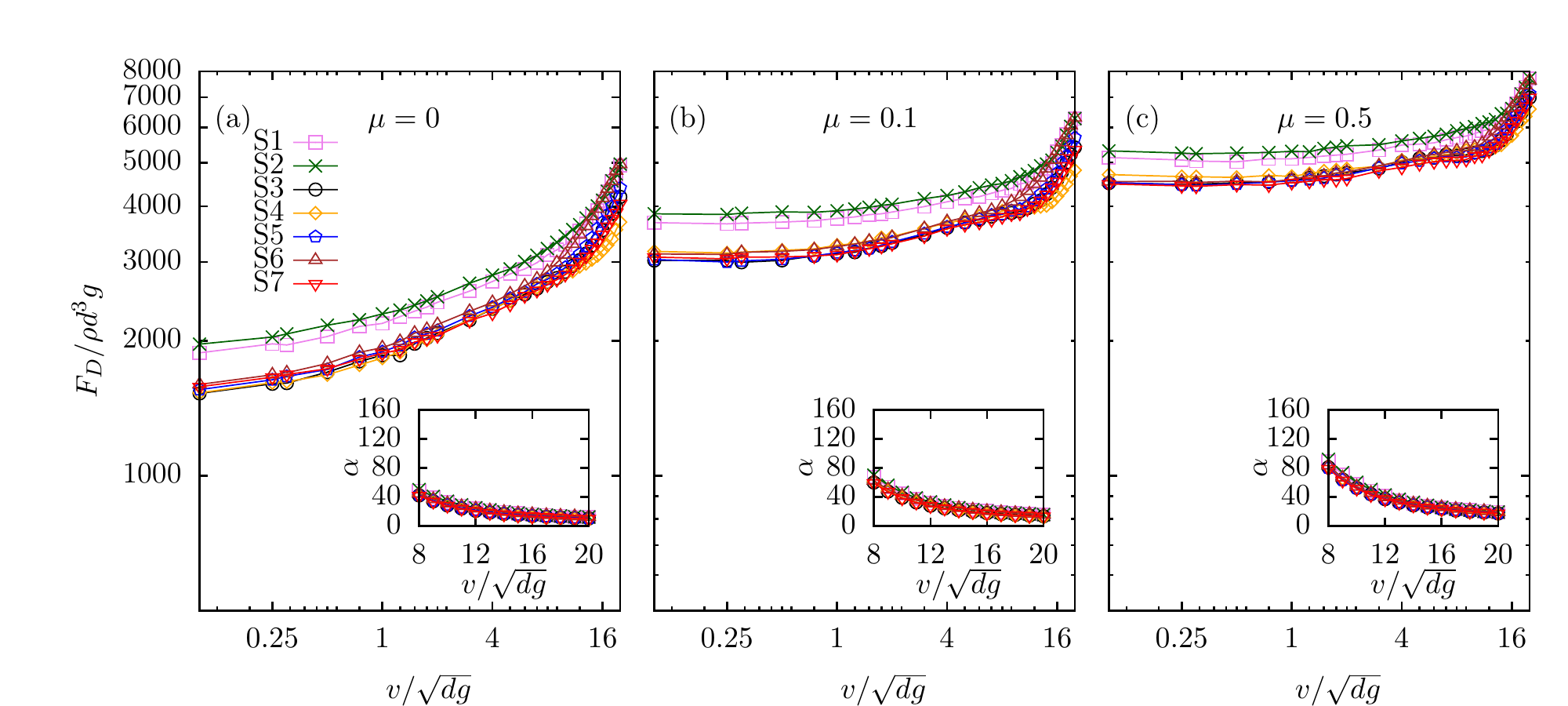}
        \caption{The variation of drag on an intruder with the velocity at (a) $\mu=0$, (b) $\mu=0.1$, and (c) $\mu =0.5$ for all the shapes in our study on a log-log scale. The inset figures show $\alpha=F_D/v^2$ for higher velocities.\label{fig:drag}}
\end{figure*}

where $m_i$ and  $d_i$ and are the mass and diameter of particle $i$,
the spring constant in the normal direction $k_n=2\times10^6\rho d g$ and the spring constant in the tangential direction $k_t=2.54\times10^6 \rho d g$. The normal and tangential damping coefficients, $\gamma_n$ and $\gamma_t$, were taken as $\gamma_n=\gamma_t=3200\sqrt{g/d^3}$. The upper limit of $\boldsymbol{f_{ij}^t}$ is restricted to $\mu\boldsymbol{f_{ij}^n}$ for considering the slipping between contacts. The coefficient of friction $\mu$ was varied between 0 to 0.5. $\boldsymbol{\hat{n}_{ij}}$ is the unit vector along the  line joining the centers of particles $i$ and $j$ and $\xi_{ij}$ is the overlap between the two particles in the normal direction. $\boldsymbol{\Delta s_{ij}}$ stands for the accumulated tangential displacement vector. Relative velocities in the normal and tangential directions are represented as $\boldsymbol{\dot{r}_{ij}^n}$ and $\boldsymbol{\dot{r}_{ij}^t}$, respectively. The model adopted is consistent with a velocity-dependent coefficient of restitution\cite{schwager98}.

The forces on the intruder are recorded at time intervals of $5\times 10^{-4}\sqrt{d/g}$. The length of the simulation corresponds to the time that it takes for an intruder to travel 1200$d$ for a specific velocity. All the simulations were performed with LAMMPS\cite{lammps} and OVITO\cite{ovito} was used for post-simulation visualisation.


\section{Results and Discussion}\label{sec:resdis}
In this section, we present the results for all seven intruder shapes (see Fig. \ref{fig:shapes}). The intruder moves at a constant velocity $v$ along the positive $x$ direction at depth $h$. The study was carried out for various velocities $v$, depth $h$, and coefficient of friction $\mu$. This section has been further split into five subsections. Subsection 3.1 presents the drag on the intruder for various velocities $v$ and coefficient of friction $\mu$ at a constant depth $h=105d$. Subsection 3.2 addresses the particle contacts on the intruder for various $v$ and $\mu$ at a constant depth $h=105d$ and kinetic drag regimes. Subsection 3.3 provides the details on  the lift force experienced by the intruder for various $v$ and $\mu$ at a constant depth $h=105d$. Subsection 3.4 explains the depth dependence of forces on the intruder for velocities $v= 1\sqrt{dg}$ and $5\sqrt{dg}$ and $\mu=0.1$. Lastly, Subsection 3.5 examines the distribution of forces around the intruder for various velocities $v$ at a fixed $h$ and $\mu$. Quantities such as drag $F_D$, lift $F_L$, and number of contacts $N_c$ reported in this work have been calculated by averaging over several configurations  after the intruder achieves a mean steady state behavior, \textit{i.e.}, the instantaneous drag fluctuates around a well defined mean.

\subsection{Drag on the moving intruder}
In this subsection, we present and discuss  the drag on the moving intruder at several velocities $v$ and coefficients of friction $\mu$. The drag $F_D$ against velocity $v$ (log-log plot) for $\mu=0,$ $0.1$ and $0.5$ is shown in Fig. \ref{fig:drag} (a), (b) and (c), respectively, for all the shapes. Additionally, $\alpha=F_D/v^2$ is shown in the insets of each figure.

A minimum force, known as the yield drag \cite{takehara,hilton,kumar}, is required to initiate the motion of an object in granular media. There is also a kinetic term associated with the drag force on the intruder\cite{kumar}. Therefore, we can write the drag as $F_D=F_Y+F_K$ where $F_Y$ is the yield drag necessary to initiate the motion, while $F_K$ is the kinetic drag. Assuming that the drag for the lowest velocity at which we performed our simulations in Fig. \ref{fig:drag} gives us the yield drag, $F_Y$ is highest for S2 followed by S1 for all $\mu$. The other shapes have nearly identical yield drag (within seven percent of each other) for all values of $\mu$ considered.

It is known from a few published studies \cite{hilton,kumar,wang,takada2016drag,takehara} that the drag regimes and drag laws in granular media are strongly correlated with the intruder's velocity. In the present work, we observe that without friction, the drag increases gradually with $v$ (see Fig. \ref{fig:drag}(a)). However, in the presence of friction, specifically, $\mu=0.1$ and $0.5$, we observe a constant drag regime at low velocities (see Fig. \ref{fig:drag}(a) and (b)). Hilton and Tordesillas\cite{hilton} and Kumar \textit{et al}\cite{kumar} studied these drag regimes in the context of a dimensionless Froude number $Fr$ which is the ratio of two timescales associated with the falling of grains in the wake and the forward motion of the intruder (assuming the intruder dimensions are much larger than the grains). They suggested that the constant drag regime exists for $Fr<1$. Applying their individual definitions of $Fr$ to our shape S3, Ref. \cite{hilton} predicts that the constant drag regime should persist until $v/\sqrt{dg}\approx1.12$, while Ref. \cite{kumar} predicts the constant drag regime to exist until $v/\sqrt{dg}\approx 1.23$.  Both definitions predict a constant drag regime in the velocity range very close to what we have observed in our study. While it is not clear how this definition of $Fr$ can be extended to non-spherical intruders, we observe that the constant drag regime occurs for almost the same velocity range for all the shapes.

Beyond the constant drag regime, the drag on the intruder increases with velocity. The slope of the curve on the log-log plot seems to be varying with $v$ suggesting that, for some constant $a$, $F_D\propto v^a$ is not valid for a granular medium. The kinetic drag contribution to the drag $F_K$ dominates yield drag $F_Y$ at high velocities. This can be observed  in the insets of Fig. \ref{fig:drag} where $\alpha=F_D/v^2$ approaches a constant value. We shall elaborate further on $F_K$ and the drag regimes in the next subsection. By comparing directly the values of $F_D$ at higher velocities ($v>10\sqrt{dg}$), we note that the drag forces are in the order: $F_{D,S1} \approx F_{D,S2}\approx F_{D,S6}>F_{D,S5}>F_{D,S3}\approx F_{D,S7}>F_{D,S4}$, where $F_{D,i}$ is the drag $F_D$ for shape $i$. Even though the differences are not large, this trend is observed consistently for all velocities higher than $10\sqrt{dg}$, and for all $\mu$ considered in our study. At velocities lower than 1.23 ($v/\sqrt(dg) < 1.23$), the trend is: $F_{D,S2} > F_{D,S1}>F_{D,S6}\approx F_{D,S5}\approx F_{D,S3}\approx F_{D,S7}\approx F_{D,S4}$.

Albert \textit{et al}\cite{albert}, in their study of the effect of shape of slowly moving object in a granular medium on jamming of grains around the object, highlighted that (a) streamlining the intruder significantly reduces the resistance offered by the granular medium to its motion, and (b) the increase in the drag on intruders that are longer in the flow direction is not due to the increase is due to the creation of a more jammed state in front of the intruder. Ding \textit{et al.} \cite{draginduced} demonstrated that the local surface stress on an intruder is approximately equal to that on a plate oriented at the same angle as the local surface and moving at the same velocity and depth. Therefore, one could calculate the forces on an intruder by summing up the individual contributions of these stresses. 

At low velocities (see Fig. \ref{fig:drag}), it is evident that streamlining an object significantly reduces the drag. An example of this would be the shape S2 (a rectangle) and S4 (an ellipse) of $20d\times10d$ along $x$ and $y$, respectively. Since an ellipse is more streamlined than a rectangle of similar dimensions, the latter has a higher drag than the former. The streamlining of a body reduces the drag since the force chains applying a force of $f_P^{fc}$ at a point $P$ on the intruder contributes only $f_P^{fc} \sin \beta$ to the drag, where $\beta$ is the tangential orientation of the intruder at point $P$ with $x$ axis ($\beta=0$ corresponds to the direction of motion). If the force chains developed were equally strong, streamlining of a body would significantly reduce drag due to only a component of force being added.
\begin{figure*}
        \includegraphics[width=0.9\linewidth]{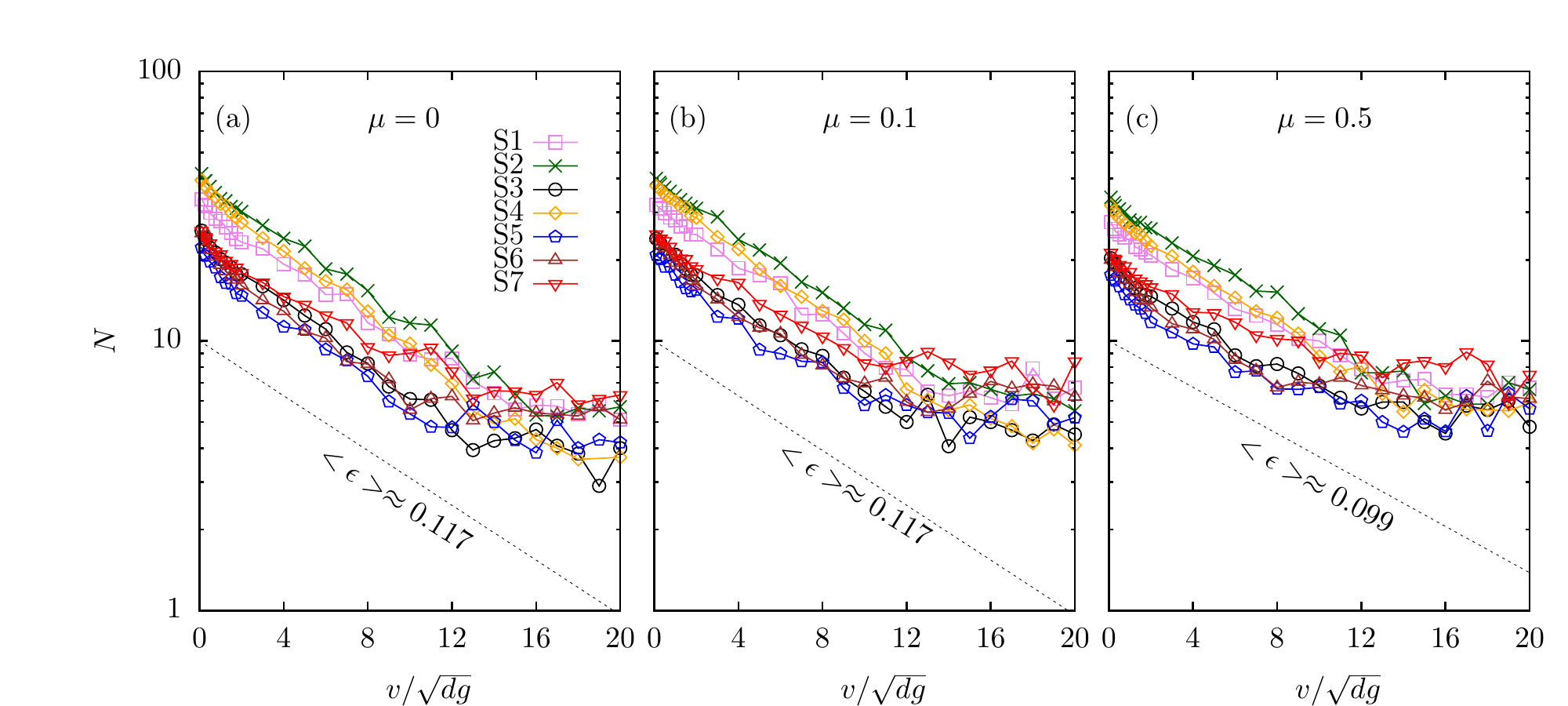}
        \caption{The variation of the mean number of contacts $N$ on an intruder with the velocity at (a) $\mu=0$, (b) $\mu=0.1$, and (c) $\mu =0.5$ for all the shapes in our study.\label{fig:contacts}}
\end{figure*}

Moreover, similar shapes such as S3, S4, and S5 have identical drag at low velocities suggesting that viscous contribution is not a major contribution to the drag force in granular media\cite{albert}. Although S4 is the longest  and it creates the most jammed force chains in front of itself compared to S3 and S5, the force chains may tend to occur more laterally, thus, contributing less to the drag. However, it must be emphasized that if the bodies are equally streamlined, the drag increases with the length of the intruder at low velocities as can be seen from $F_D$ for shapes S1 and S2. This is because longer shapes (Fig. 3) delay the collapse of force chains in the intruder's wake allowing those with higher stress to be formed in front of the intruder. Also since the forward facing surfaces of S1, S2 and S6 are blunt ($\sin \beta=1$), the force chains contribute more to drag compared to the other shapes. Moreover, the shape S6 allows the force chains to collapse almost immediately and thus, has lower drag than S1 or S2. S7 is also streamlined and, therefore, has less drag than S1 or S2 at lower velocities. This will be shown in later subsections.

As previously stated, at higher velocities, the drag forces are in the order $F_{D,S1} \approx F_{D,S2}\approx F_{D,S6}>F_{D,S5}>F_{D,S3}\approx F_{D,S7}>F_{D,S4}$. Firstly, S1, S2 and S6 are subject to an approximately equal drag, unlike at low velocities. Interestingly, Ding \textit{et al}\cite{draginduced} showed that stress acting on a flat plate being dragged in granular media is identical to the small element on the intruder with a similar orientation to the flow $\beta$, irrespective of the shape of the intruder. Since the front faces (side facing the flow) of these three intruders are oriented at the same angle to the flow, they experience the same magnitude of stress on the front face. Additionally, the other faces contribute very little to the drag as they do not experience contacts with the grains.  This leads to the three shapes S1, S2 and S6 having an identical drag. It is noteworthy that, at low velocities, the drag is not equal for the three shapes, because of the way these shapes hinder the jamming and collapse of grains around them.

Other shapes with curved front surfaces experience a lower drag than S1, S2, and S6. The curved surfaces contribute $\sin \beta<1$ times the normal force on the intruder's surface to the drag. Therefore, the more blunt an object is, the more drag it experiences. Hence, S5 has a lower drag than S1, S2, and S6, but higher than S3. As the shape becomes less blunt, the drag is further reduced with S4 having a lower drag than S3. Interestingly, S7 also has a higher $F_D$ than S4 because two of the sides of S7 are tilted at an angle of $\pi/3$ to the flow. This is very close to  the angle at which a plate moved through granular media experiences the maximum stress\cite {draginduced}. S3 and S7 experience similar drag due to the summation of drag components being almost similar. We discuss the force profiles on the intruders in  Sec. \ref{sec:prof}.

\subsection{Number of contacts, kinetic drag, and drag regimes}
In order to understand the forces acting on an intruder, it is imperative to reflect on the role of the grains in contact with the intruder: it is these grains that are `directly' responsible for the forces. The average number of grain contacts $N$ changes with the intruder's shape as well as its velocity. Therefore, in this subsection, we present the results and discussion on $N$ and the relation between $N$, $F_K$ and the drag regimes.
\begin{table}[h!]
        \centering
        \setlength{\tabcolsep}{12pt}
        \caption{Comparison of $\phi$ obtained from numerical simulations and the theoretical estimate, $l_{cs}/S$. \label{table:table2}}
        \begin{tabular}{c c c}
                \hline
                \hline
                \rule{0pt}{2.5ex}
                Shape   &       $\phi_\textrm{simulations}$     &       $\phi=l_{cs}/S$\\
                \hline
                S1    & 0.20 & 0.25\\
                S2    & 0.24 & 0.20 \\
                S3    & 0.22 & 0.32\\
                S4    & 0.17 & 0.21\\
                S5    & 0.27 & 0.37\\
                S6    & 0.30 & 0.33\\
                S7    & 0.34 & 0.33\\
                \hline\\
        \end{tabular}
\end{table}

The number of contacts $N$ is calculated by averaging over several configurations after the intruder has reached a steady-state behavior, \textit{i.e.}, when the drag force fluctuates around a well-defined mean. Of course, $N$ is a strong function of the shape of the intruder, but it does not imply that two shapes with same number of contacts at a given $v$ experience equal forces. For example, a blunt or streamlined object may a have similar number of contacts.

The variation of $N$ with $v$ is shown in Fig. \ref{fig:contacts} (a), (b), and (c) on a semi-log plot for $\mu=0,$ 0.1, and 0.5, respectively. Two distinct regimes can be identified; an exponential one in which $N\propto e^{\epsilon v}$ and a second in which $N=N_{\infty}$ is constant. The latter seems to be in the same velocity range in which the $v^2$ dependence of drag force is expected. We do not exactly understand the physical picture behind this saturation, but it is consistently present for all the shapes and for all the $\mu$ considered in our study. However, this saturation value can be approximated as $N_{\infty}\approx\Phi l_{cs}/d$, where $\Phi$ is the packing fraction of the bed and $l_{cs}$ is the cross-section length perpendicular to the direction of motion. Let $\phi=N_{\infty}/N_0$ where $N_0$ is the number of contacts in the yield limit, which is roughly equal to $\Phi S/d$ where $S$ is the perimeter of the obstacle. Therefore, $\phi\approx l_{cs}/S$. We have compared this rough theoretical estimate of $\phi$ with that obtained from the numerical simulations in Table \ref{table:table2}. Additionally, we observe that $\epsilon\approx -0.1$ as can be seen in Fig. \ref{fig:contacts} (a), (b) and (c) and the value of $\epsilon$ for each individual shape and $\mu$ is presented in Table \ref{table:table1}.
\begin{table*}
        \centering
        \setlength{\tabcolsep}{18pt}
        \caption{The values of fits and constants}        \begin{tabular}{c c c c c c c}                \hline                \hline                \rule{0pt}{2.5ex}
                Shape   &       $\mu$   &       $-\epsilon$     &       $\zeta_1$ & $R^2_{\zeta_1}$ & $\zeta_2$ & $R^2_{\zeta_2}$\\
                \hline\\
                \multirow{ 3}{*}{S1} & 0.0 & 0.133 & 1.064       &       0.981   &       1.970   &       0.983\\
                & 0.1 & 0.120 & 1.319   &       0.993   &       2.168   &       0.973\\
                & 0.5 & 0.099 & 1.296   &       0.999   &       2.411   &       0.976\\
                \\  
                \multirow{ 3}{*}{S2} & 0.0 & 0.115 & 1.106       &       0.984   &       2.244   &       0.974\\
                & 0.1 & 0.119 & 1.403   &       0.997   &       2.591   &       0.991\\
                & 0.5 & 0.100 & 1.440   &       0.961   &       2.420   &       0.986\\
                \\
                \multirow{ 3}{*}{S3} & 0.0 & 0.126 & 1.130       &       0.986   &       1.752   &       0.922\\
                & 0.1 & 0.123 & 1.290   &       0.997   &       1.938   &       0.942\\
                & 0.5 & 0.099 & 1.186   &       0.991   &       2.094   &       0.955\\
                \\
                \multirow{ 3}{*}{S4} & 0.0 & 0.132 & 1.126       &       0.993   &       2.143   &       0.988\\
                & 0.1 & 0.133 & 1.301   &       0.997   &       2.300   &       0.978\\
                & 0.5 & 0.116 & 1.363   &       0.991   &       2.188   &       0.975\\
                \\
                \multirow{ 3}{*}{S5} & 0.0 & 0.122 & 1.063       &       0.983   &       1.422   &       0.905\\
                & 0.1 & 0.113 & 1.25    &       0.994   &       1.746   &       0.945\\
                & 0.5 & 0.090 & 1.191   &       0.979   &       2.022   &       0.945\\
                \\
                \multirow{ 3}{*}{S6} & 0.0 & 0.122 & 1.106       &       0.981   &       1.296   &       0.966\\
                & 0.1 & 0.113 & 1.327   &       0.998   &       1.601   &       0.951\\
                & 0.5 & 0.094 & 1.296   &       0.993   &       2.08    &       0.976\\
                \\
                \multirow{ 3}{*}{S7} & 0.0 & 0.092 & 1.001       &       0.994   &       1.579   &       0.951 \\
                & 0.1 & 0.099 & 1.292   &       0.997   &       1.807   &       0.911 \\
                & 0.5 & 0.073 & 1.225   &       0.983   &       2.147   &       0.932 \\
                \\
                \multirow{ 3}{*}{Mean}   & 0.0 & - & 1.001       &   -   &       1.579   & - \\
                & 0.1 & - & 1.292       &       -       &       1.807   & - \\
                & 0.5 & - & 1.225       &       -       &       2.147   & - \\
                \\
                \hline
                \hline
        \end{tabular}
        \label{table:table1}
\end{table*}

Based on plots of $F_D$ and $N$ versus $v$ for $\mu\ge0$ and by comparing their respective behaviors for all the intruder shapes considered in the present study, we propose a three-regime model for the average drag force $F_D$ acting on the intruder. In the first, which is observed in the range $1<v/\sqrt{dg}<4$, the drag is constant; the second occurs in the range $8<v/\sqrt{dg}<12$ and the third for  $10<v/\sqrt{dg}<12$. In the remainder of this subsection, we explore the dependence of the kinetic drag on the velocity for each of these regimes.

Each contact exerts a force on the intruder whose magnitude depends on its location with respect to the intruder's line of motion. It is difficult to correlate the average number of contacts $N$ with the force on the intruder. A previous study \cite{hilton} proposed $F_K\propto N v^\zeta$ with $\zeta=1$, but they considered only the first two drag regimes. Here we assume that $\zeta$ can vary depending on the regime:

\begin{align}
\textrm{Regime I:} & \quad\quad F_K=0\\
\textrm{Regime II:} & \quad \quad F_K=\alpha v^{\zeta_1} N/N_0 =\alpha \exp(\epsilon v) v^{\zeta_1}\\
\textrm{Regime II:} & \quad \quad F_K=\beta v^{\zeta_2} N/N_0 =\beta  v^{\zeta_2} l_{cs}/S
\end{align}

Best-fit parameters of these equations to our simulation data, along with their coefficient of determination, are presented in Table \ref{table:table1}. In regime II, it can be seen that $\zeta_1\approx 1$ for all shapes in accord with the previous results \cite{hilton} for $\mu>0$. As for the regime III, $\zeta_2\approx 2$ is consistent with a $v^2$ dependence of kinetic drag with velocity. Except for the frictionless ($\mu=0$) systems where the trends are not clear, the proposed regimes work well for all the shapes with $\zeta_1\approx 1 $ and $\zeta_2 \approx 2$.


\subsection{Lift on the moving intruder}

\begin{figure*}
        \includegraphics[width=0.9\linewidth]{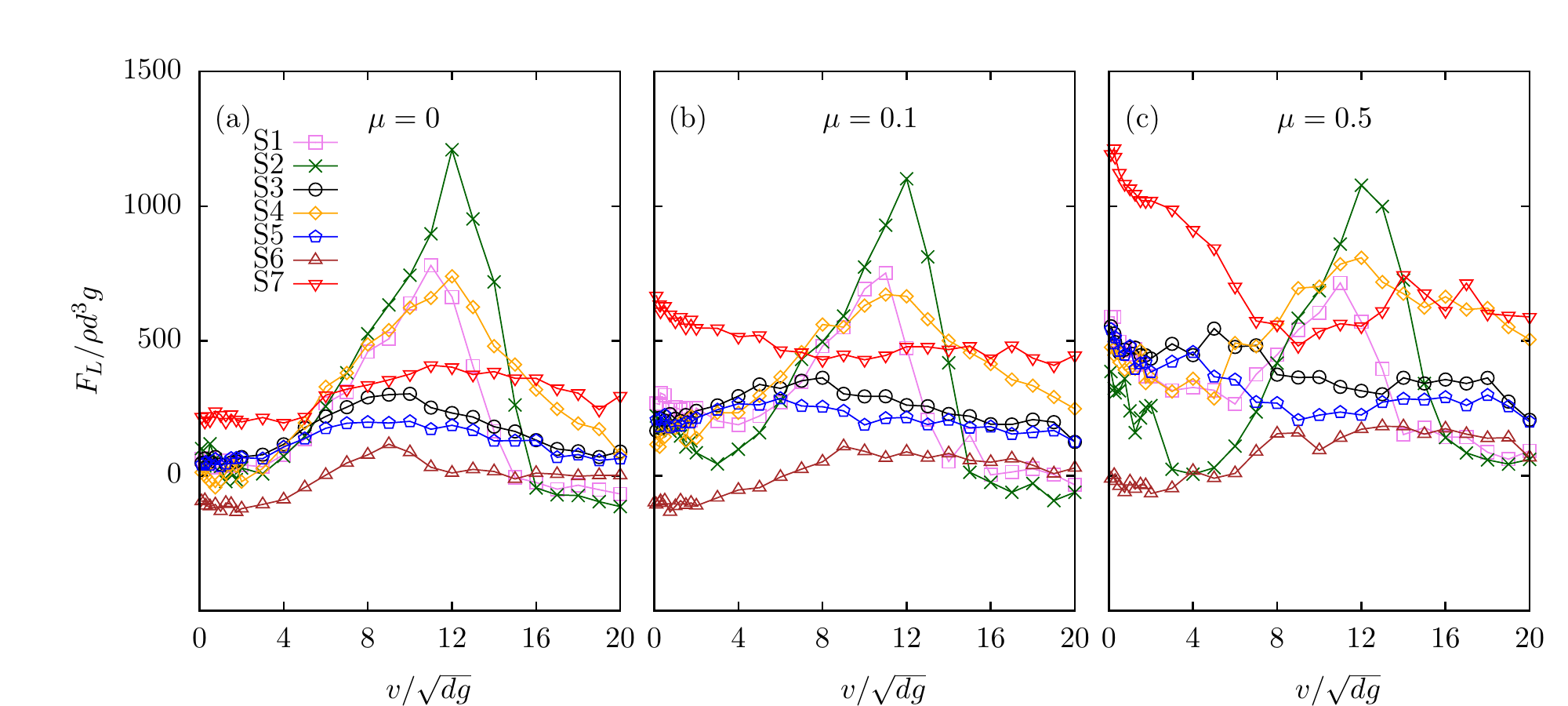}
        \caption{The variation of lift on an intruder with the velocity at (a) $\mu=0$, (b) $\mu=0.1$, and (c) $\mu =0.5$ for all the shapes in our study.}\label{fig:liftforce}
\end{figure*}
\begin{figure}[h]
        \includegraphics[width=1\linewidth]{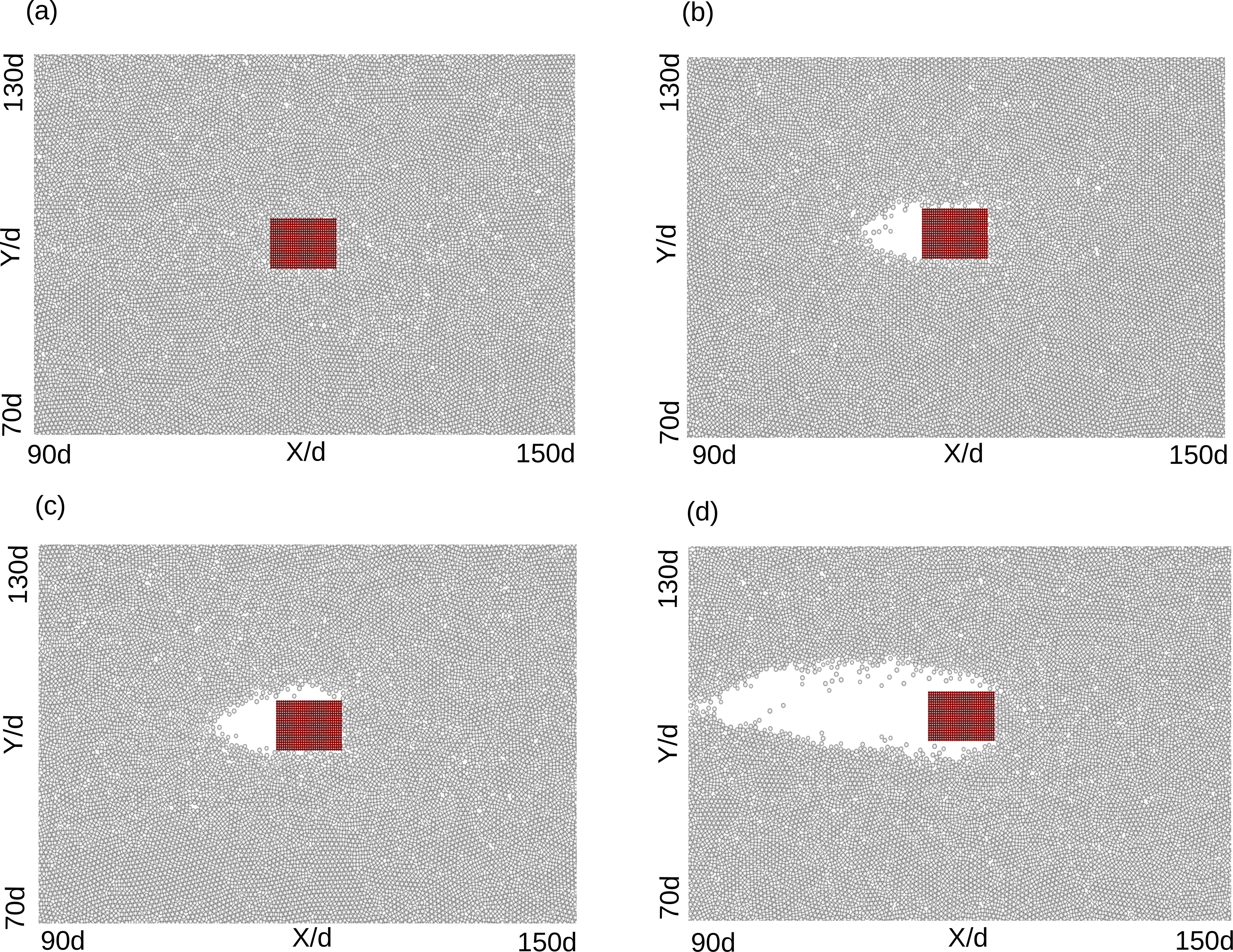}
        \caption{Snapshots showing the region around the $S2$(rectangle) intruder for $\mu=0$ at different intruder velocities ($v/\sqrt{dg}$) (a)  1, (b) 8, (c) 12 and (d) 18.\label{fig:s2snap} }
\end{figure}

        The  component of the force on the intruder perpendicular to the flow direction is defined as the lift force, $F_L$. Figure \ref{fig:liftforce} shows the  variation in $F_L$ as a function of the intruder's velocity ($v$) for various coefficients of friction ($\mu$ = 0, 0.1 and 0.5). For a clear exposition of the trends, we consider three velocity regimes: (i) low velocities ($0.1<v/\sqrt{dg}<4$); (ii) intermediate velocities ($6<v/\sqrt{dg}<14$); and (iii) high velocities ($14<v/\sqrt{dg}<20$). For $\mu=0.0$ (Fig. \ref{fig:liftforce}(a)), a maximum in $F_L$ for all intruder shapes is observed in the intermediate velocity regime and is ordered as follows: $F_{L,S2} > F_{L,S1}> F_{L,S4}>F_{L,S7}>F_{L,S3}>F_{D,S5}>F_{L,S6}$. For most of the shapes $F_L$ saturates in the third regime. The highest $F_L$ is observed for $S2$ (rectangle) in the intermediate regime. In both frictional and frictionless systems the maximum lift force for this shape is observed at $v/\sqrt{dg}=12$ followed by a sharp decrease. The non-monotonic behavior of the lift force is result of flow detachment from the intruder. One such example is shown in Fig. \ref{fig:s2snap} depicting simulation snapshots of the flow region around the $S2$ shape for $\mu = 0.0$ at different velocities. For $v/\sqrt{dg}=1$  the granular particles just slide past the intruder without leaving a trail behind it. At $v/\sqrt{dg}=8$  flow detachment from the upper surface of  the intruder is evident. This can be attributed to the presence of a free surface at the top and a confined wall at the bottom of the assembly of particles. As a result, the particles below the intruder exert a net upward force. As $v$ increases, flow detachment continues to occur only from above the intruder but not below it. The expansion of flow detachment above the intruder is observed until $v/\sqrt{dg}=12$ .  Beyond $v/\sqrt{dg}=12$, however, the intruder imparts more energy to the granular particles in its path, which eventually results in flow detachment from both the upper and lower surfaces of the intruder as shown in Fig. \ref{fig:s2snap} (d) for $v/\sqrt{dg}=18$.

 It is generally observed that above a certain velocity particle contact only occurs on the leading side of the intruder with flow detachment from both its top and bottom surfaces. This flow detachment plays an important role in the decrease of $F_L$ beyond a certain $v$ for shapes $S1$ (square), $S2$ (rectangle) and $S4$ (ellipse with the major axis aligned with the $x$-direction), compared to the other intruders. The lift force on  the equilateral triangle $S7$ (with edge pointing to the moving direction), disc $S3$, and ellipse major $S5$ shapes varies little with $v$ for $\mu=0.0$. This could be due to the maximum particle interaction being concentrated on the frontal part. Interestingly, a negative $F_L$ is observed for an equilateral triangle with its edge pointing opposite to its direction of motion ($S6$) in the low velocity regime. In this case, particles traversing the edges of the blunt surface exert more force on the upper inclined surface than that on the lower one. This is because the particles fall on the upper inclined surface of $S6$, while the particles have to move against gravity to reach the lower inclined surface.  It is also evident  that the lift force saturates for most of the shapes in higher velocity regime.

 \begin{figure*}[ht!]
        \includegraphics[width=0.9\linewidth]{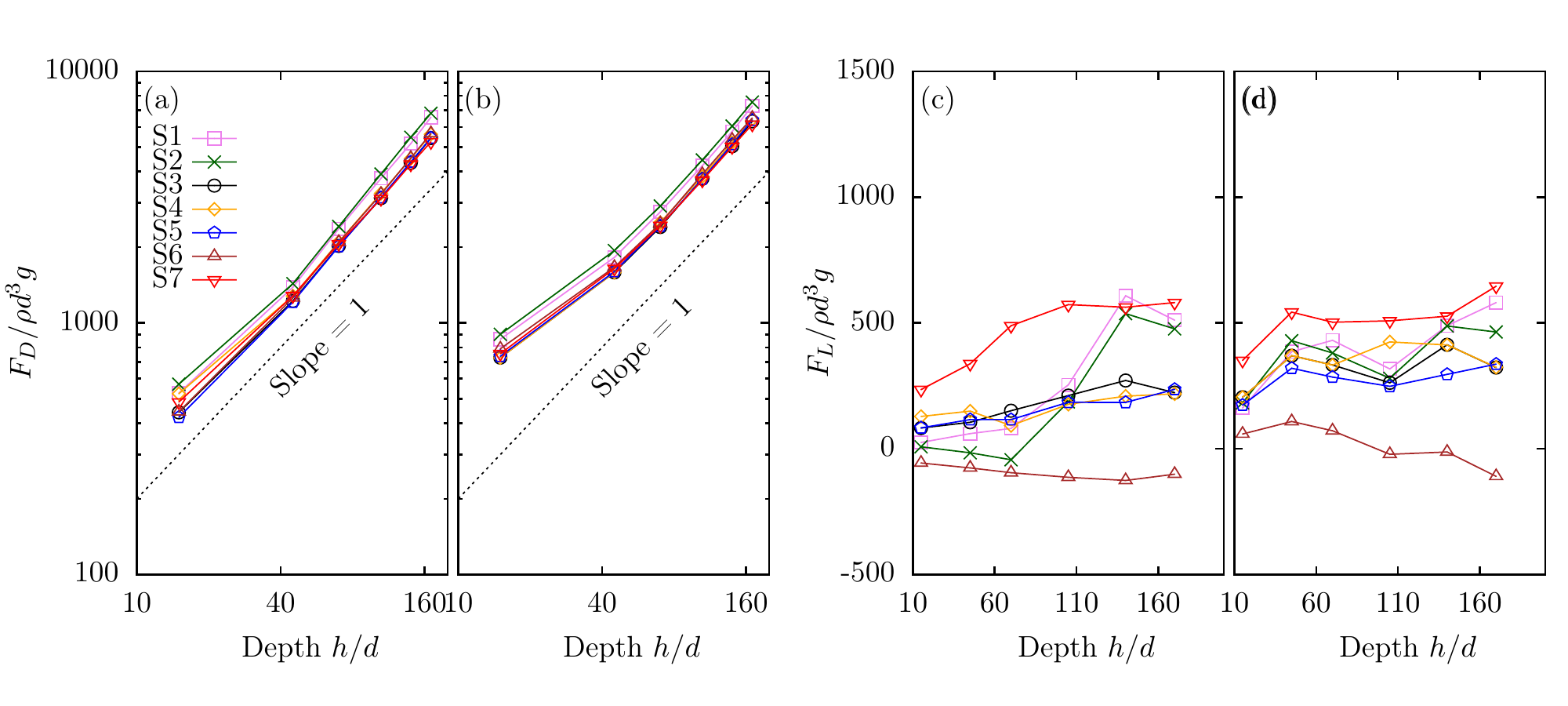}\\
        \caption{The variation of drag force on an intruder at $\mu=0.1$ with depth for (a) $v/\sqrt{dg}=1$ and (b) $v/\sqrt{dg}=5$ on a log-log scale. The dotted line represents a straight line of slope 1 on the log-log scale. The variation of lift force on an intruder at $\mu=0.1$ with depth for (c) $v/\sqrt{dg}=1$ and (d) $v/\sqrt{dg}=5$ on a linear scale. \label{fig:depth}}
 \end{figure*}

  For systems with friction coefficients $\mu=0.1$ and $0.5$, the lift force on $S1, S2$ and $S4$ is  higher than on the other shapes in the intermediate velocity regime. The reason being the flow detachment and the larger contact surface around the top and bottom region of the intruder. The lift force on shapes $S3, S5,$ and  $S6$  show little variation for $\mu=0.1$ for a range of $v$, while for $\mu=0.5$ the lift force is higher for $S3$ and $S5,$ in the low velocity regime and then gradually saturates in the high velocity regime. Unlike the other shapes the net lift force on $S6$ is small for all values of $\mu$ and is negative in the low velocity regime.

  The two equilateral triangles $S6$ and $S7$ have different orientations (edges pointing opposite and along the moving direction, respectively). This significantly impacts the $F_L$ experienced by the two shapes as seen in Fig. 5. Ding \textit{et al.} have also stated that $F_L$ is sensitive to the cross-section of the intruder. The $S7$ has higher $F_L$ than the $S6$ for all the friction coefficients. The particles in front of the intruder collide with a flat surface in the case of $S6$, whereas they collide with a "V" shaped surface with its edge pointing in the direction of the intruder's motion for the case of $S7$. For $\mu=0.1$ the lift force was maximum at low velocity regime for the $S7$ with gradually saturating at higher velocity(Fig. 5(b)). The $S7$ shape has an pointed edge at the moving direction of intruder with its two frontal surfaces inclined to each other at $60^{\circ}$ angle and a blunt back. It does have a larger area for particle contact at the leading surface. At low velocities the particles coming in contact at the upper inclined surface slides past the intruder. While there is an accumulation of particles at the bottom front surface due to the confined wall at the bottom. Thus giving it a push in the positive y-direction. This can be a reason that $S7$ is having higher $F_L$ at low velocity regime.

  Another interesting observation is the  independence of $F_L$ on the intruder's velocity in the range $0.1<v/\sqrt{dg}<2$ for $\mu=0.0$ for all the shapes. This behavior is in contrast with the trend of drag force, where a gradual increase is observed with intruder velocity. For both frictional and frictionless systems,  the maximum $F_L$ is observed for $S2$(rectangle) in the intermediate velocity regime. The lift force has a dependence on the orientation of the intruder, for example, $S6$ and $S7$  exhibited completely different behavior to each other.


\subsection{Depth-dependence of forces on the moving intruder}

 Albert \textit{et al}\cite{albert} observed a nonlinear depth dependence of the drag force on a discrete object moving at a very low velocity immersed in a granular bed at a depth of 40 - 150 $mm$ (equivalent to 44 - 166 granular bed particles depth). Guillard \textit{et al.}\cite{guillardprl} observed a depth-independent drag force on a cylindrical object immersed deep (120 particle lengths or more) in a granular medium and rotated about the vertical axis. A few studies \cite{potiguar2013lift,PhysRevE.91.022201,panaitescu2017drag} also reported a linear depth dependence. In the context of these results, we compare $F_D$ on the different intruder shapes as a function of their immersion depth in Fig. 7 (a) and (b). The results are shown for two intruder velocities $v/\sqrt{dg}$= 1 and 5 at $\mu=0.1$. In the simulations, the intruder is placed at six different depths $h/d = 15, 45, 70, 105, 140, 170$. $F_D$ increases linearly with an increase in $h/d$ for all the intruders considered here, confirming that the drag is proportional to the hydrostatic pressure. In granular medium the number of particles above the intruder increases with an increase in its depth thus increasing hydrostatic pressure. The $F_D$ for the various h/d at $v/\sqrt{dg}$=1 and $\mu=0.1$ can be ordered as: $F_{D,S2} > F_{D,S1}>F_{D,S4}\approx F_{D,S7}\approx F_{D,S3}\approx F_{D,S5}\approx F_{D,S6}$. The same trend is also observed for $v/\sqrt{dg}=5$. The $S1$ and $S2$ experiences the maximum drag for all $h/d$ for the two $v$ shown in the plots while the other shapes have identical $F_D$ within five percent of each other at a particular depth. Moreover, the difference between the highest ($S2$) and lowest ($S7$) drag forces calculated for one random depth is not more than $25\%$.

 It has been reported in the literature that the lift force either saturates  \cite{stress3,debnath2017lift} or increases \cite{draginduced} with the immersion depth of intruder in a granular medium. To determine the shape and depth dependence, we plot the lift force $F_{L}$ as a function of $h/d$ for two intruder velocities, $v/\sqrt{dg}$=1 and 5, at $\mu=0.1$  in Fig. \ref{fig:depth} (c) and (d). Minimal change in  lift force  for intruders $S3$, $S4$ and $S5$ is observed  at $v/\sqrt{dg}=1$. The lift force on intruders $S1$ and $S2$  increases sharply below a certain depth for $v/\sqrt{dg}=1$. For  the same velocity $F_L$ on $S7$  increases up to a certain depth and then saturates with further increase in the depth. For $v/\sqrt{dg}=5$ there is a fluctuation in lift force for all the intruders except for $S6$ which shows a gradual decrease in $F_L$ with an increase in $h/d$. $S6$ and $S7$ with the same geometry but different orientation in the $x$-direction have the lowest and highest lift forces, respectively for  almost at all the depths considered. The lift force acting on $S6$  moving at $v/\sqrt{dg}$=1 shows little variation with depth, while at $v/\sqrt{dg}=5$ it decreases with the intruder depth.

While the results of Fig. \ref{fig:depth} suggest that the relation between drag and depth can be easily understood the same is not true for the lift force, even though  both forces result from the repulsive interactions between granular particles and the intruder. The lift force experienced by the intruder also depends upon the number of particle contacts at its upper and lower surface. So the forces depend on the specific region of contact around the intruder surface. Thus if the sum of forces acting perpendicularly on the lower surface of the intruder is more due to the particle contacts, then it experiences a positive $F_L$ and otherwise it has a negative $F_L$. In the next section, we examine the force distribution on the intruder surface and how it influences the drag and lift forces.

 \begin{figure}[h]
        \includegraphics[width=1\linewidth]{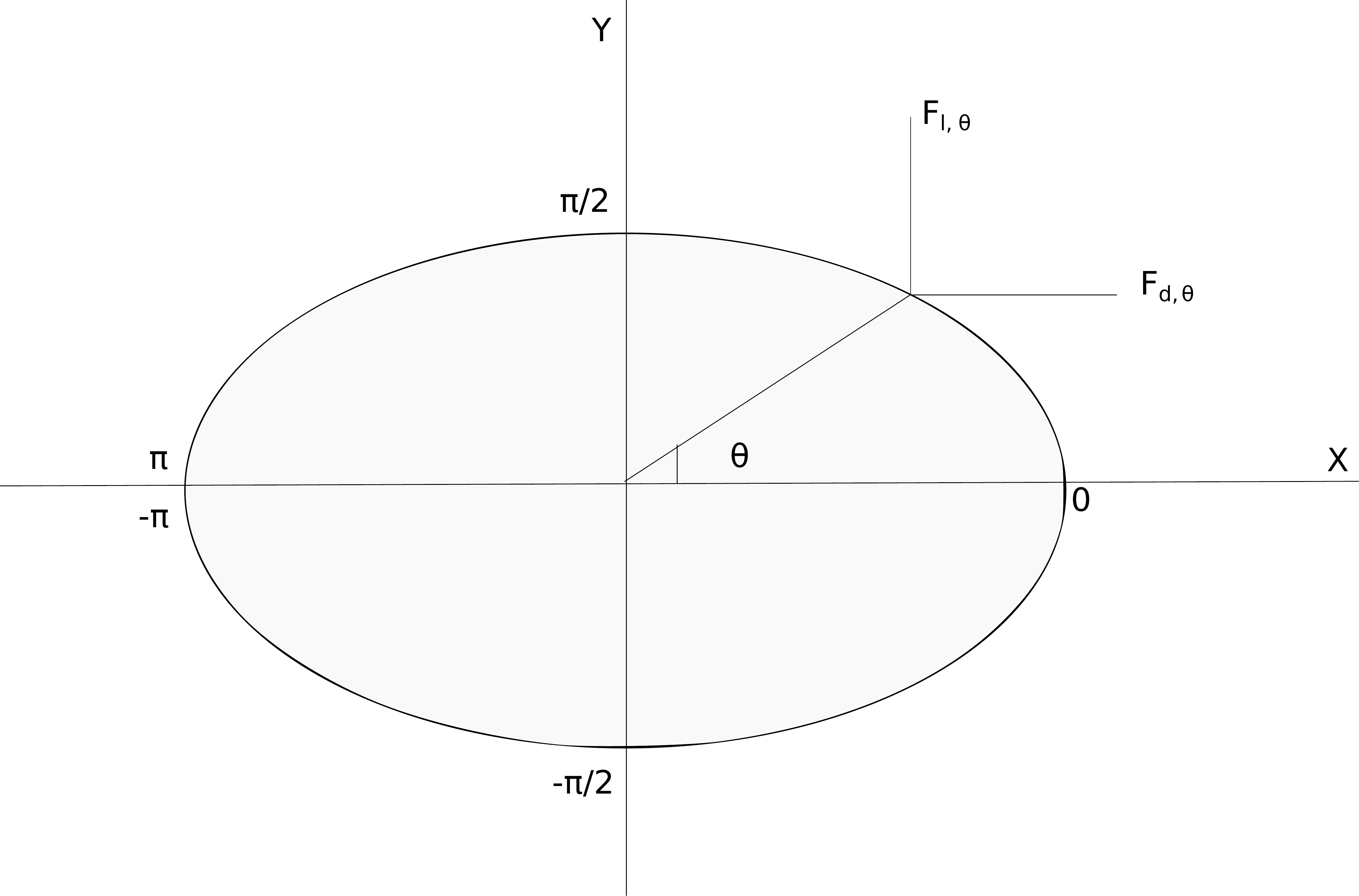}

        \caption{Geometric representation of an $S4$ intruder (an ellipse with the major axis aligned with the direction of motion of the intruder). The contact position is given by the angle $\theta$ with $-\pi<\theta<\pi$ with the upper (lower) half corresponding to positive (negative) values.\label{fig:theta}}
 \end{figure}

 \begin{figure*}
        \subfloat[][]{\includegraphics[clip,trim=0cm 0.8cm 0.2cm 0.2cm, width=0.45\linewidth]{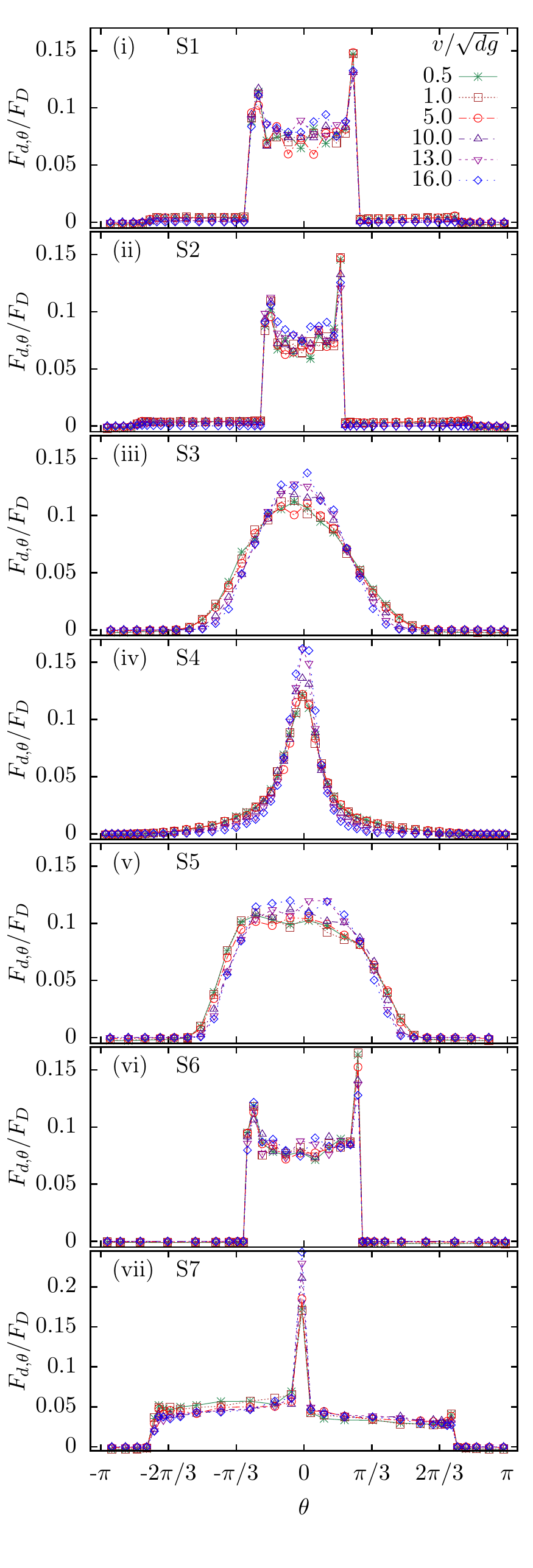}}
        \subfloat[][]{\includegraphics[clip,trim=0cm 0.8cm 0.2cm 0.2cm, width=0.45\linewidth]{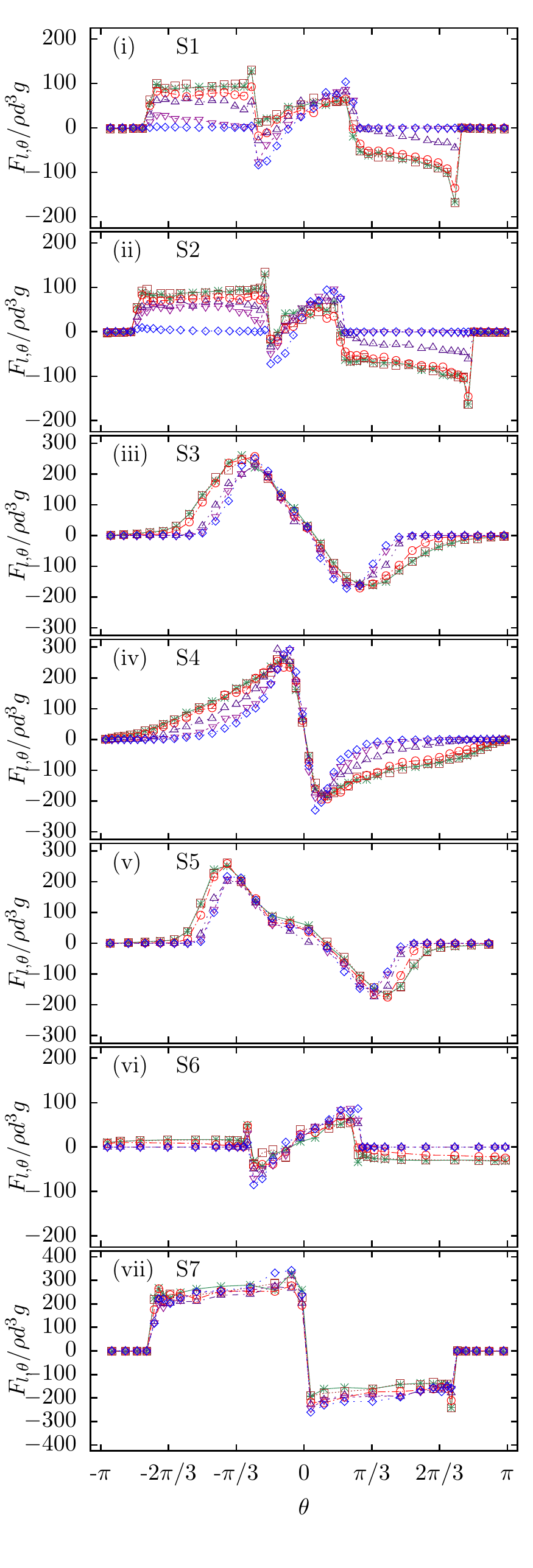}}
        \caption{The variation of (a) the normalized drag force ($F_{D,\theta}$/$F_D$) and (b) lift force ($F_{l,\theta}$) around the intruder surface for various intruder velocities. Please refer to Fig. \ref{fig:theta} for the definition of $\theta$. \label{fig:forcedist}}
\end{figure*}


\subsection{Force profile along the surface of the intruder \label{sec:prof}}

To develop a better understanding of how the forces act around the periphery of a moving intruder, the force distribution as a function of the angle of contact ($\theta$),(definition of $\theta$ in Fig. \ref{fig:theta}) is analyzed. Positions on the upper (lower) surfaces correspond to positive (negative) angles. The number of particles in contact is more at the front of the intruder with an increase in $v$ while on the backside there is a decrease as shown in Fig. \ref{fig:forcedist}. The shapes $S1$, $S2$ and $S6$, having a blunt face to the moving direction has higher drag for $\theta$ value, $\pi/3$ to $-\pi/3$. This is due to the maximum particle contact being concentrated at the face of the intruder in the moving direction. It is also observed that there is a sharp increase in drag at the leading edges. This is because when the blunt-faced intruders are moving within the medium, they push the particles along with it. These colliding particles will have an opposite perpendicular force acting on the blunt face of the intruder whereas at the leading edges the force with which the particles interact will be more due to the gravitational force that comes into play. There are more particle contacts at the backside of the intruder at low velocities compared with the case of  high velocities(no visible contacts as shown in Fig. 6). The empty wake created at high $v$ behind the intruder  is due to the detachment of particles around the obstacles at high drag speeds \cite{potiguar2013lift}. The drag force experienced by the intruder is mainly due to its contacts with particles at the front in the direction of motion of the intruder. At the back of the intruder surface, the particles exert minimal contact forces as they just slide down due to gravity after  contacting the leading surface of the intruder. Thus the force is almost zero for $\pi/2<|\theta|<\pi$. The curved intruders $S3$ and $S5$ have a maximum drag force for $\theta=0$. The shapes $S4$ and $S7$  which have a pointed edge to its  direction of motion, experiences a higher drag at $\theta=0$. The $S7$ shape has maximum drag force at its pointed edges  and then followed by a sharp decrease at the sides of the intruder tilted at an angle of $\pi/3$ to the moving direction.

 Fig. \ref{fig:forcedist} (b) shows the distribution of lift force around the intruder surface for various $v$. The shapes $S1$, $S2$ and $S6$ exhibit similar lift force profiles for $|\theta|<\pi/4$ as they present a blunt face in the leading direction. $S2$ (rectangle) experiences  high lift force for $\pi/4< \theta < 2\pi/3$. The curved intruders $S3, S4$ and $S5$ exhibit a higher positive lift force suggesting a larger number of particle contacts on their lower surface. This is due to their more streamlined form compared to other shapes. The equilateral triangles $S6$ and $S7$ show different lift force profiles. The former, with its blunt front face, experiences essentially zero lift on its trailing sides ($|\theta|>\pi/3$), while the $S7$ experiences strong positive lift forces on its lower leading face  $-2\pi/3<\theta<0$ and a weaker negative lift force on its upper leading face, $0<\theta<2\pi/3$.
\\

\section{Conclusions}\label{sec:conclu}
We have presented numerical simulation results of an intruder dragged horizontally through a granular medium to understand the drag and lift forces it experiences as a function of its velocity ($v$), immersion depth ($h/d$) and shape. The drag force gradually increases with $v$ in frictionless systems ($\mu=0.0$), while when friction is present we observe a constant drag regime at low velocities. For a fixed cross section, the drag force depends  weakly on the intruder shape. In contrast, the lift force has a strong shape dependence. It may increase  in a certain velocity range but at higher velocities a decline in the lift force is noticed. The intruder shape has a major effect on the distribution of contacts around its surface, which explains the strong lift experienced by certain shapes. The force profiles around the intruder surface resulting from granular contacts exhibit a strong angular dependence.

\section*{Conflicts of interest}
There are no conflicts to declare.

\section*{Acknowledgements}
We thank Param-Ishan, computing facility at IIT Guwahati.


\section*{References}

\bibliography{mybibfile} 
\end{document}